\documentclass[oldversion]{aa}  
\usepackage{graphicx}
\usepackage{txfonts}

\def\pro2{\textsc{\bfseries pro2}}
\def\lte2{\textsc{\bfseries lte2}}
\def\atoms2{\textsc{\bfseries atoms2}}
\def\setf2{\textsc{\bfseries setf2}}
\def\line1prof{\mbox{\textsc{\bfseries line1}\raisebox{-0.5ex}{\bfseries--}\textsc{\bfseries prof}}}

\def\etal{{et\,al.}\ }

\def\keins{K1-16}
\def\hh{H1504$+$65}
\def\rxj{RX\,J2117.1$+$3412}
\def\ngc{NGC\,246}
\def\lovier{Longmore\,4}
\def\kpd{KPD\,0005+5106}
\def\sand{Sand~3}
\newcommand{\Teff}{$T\mathrm{\hspace*{-0.4ex}_{eff}}$\,}
\newcommand{\logg}{$\log\,g$\hspace*{0.5ex}}

\newcommand{\gppr}{\stackrel{>}{\scriptstyle \sim}}
\newcommand{\gappr}{\raisebox{-0.4ex}{$\gppr $}}

\begin{document}
\title{Identification of \ion{Ne}{viii} lines in  H-deficient (pre-)
  white dwarfs: a new tool to constrain the temperature of the hottest
  stars\thanks{Based on observations made with the NASA-CNES-CSA Far
  Ultraviolet Spectroscopic  Explorer. FUSE is operated for NASA by the
  Johns Hopkins University under NASA contract NAS5-32985.}  \thanks{
  Some of the data presented in this paper were obtained from the
  Multimission Archive at the Space Telescope Science Institute
  (MAST). STScI is operated by the Association of Universities for
  Research in Astronomy, Inc., under NASA contract NAS5-26555. Support
  for MAST for non-HST data is provided by the NASA Office of Space
  Science via grant NAG5-7584 and by other grants and contracts.} }

\author{K\@. Werner\inst{1}
   \and T\@. Rauch\inst{1}
   \and J.~W\@. Kruk\inst{2}}

\institute{Kepler Center for Astro and Particle Physics, 
Institut f\"ur Astronomie und Astrophysik, Universit\"at T\"ubingen, Sand 1, 72076 T\"ubingen, Germany
   \and Department of Physics and Astronomy, Johns Hopkins University, Baltimore, MD 21218, USA}

\offprints{K\@. Werner\\ \email{werner@astro.uni-tuebingen.de}}

\date{Received; accepted}

\authorrunning{K. Werner \etal}
\titlerunning{\ion{Ne}{viii} in the hottest hydrogen-deficient (pre-) white dwarfs}

\abstract{For the first time, we have identified  \ion{Ne}{viii}
absorption lines in far-UV spectra of the hottest known
(\Teff$\gappr$150\,000~K) hydrogen-deficient (pre-) white dwarfs of
spectral type PG1159. They are of photospheric origin and can be matched
by synthetic non-LTE line profiles. We also show that a number of UV and
optical emission lines in these stars can be explained as being
photospheric \ion{Ne}{viii} features and not, as hitherto suspected, as
ultrahigh ionised \ion{O}{viii} lines created along shock-zones in the
stellar wind. Consequently, we argue that the long-standing
identification of the same emission lines in hot [WR]-type central stars
as being due to ultrahigh-ionised species (\ion{O}{vii-viii}, \ion{C}{v-vi})
must be revised. These lines can be entirely attributed to thermally excited
species (\ion{Ne}{vii-viii}, \ion{N}{v}, \ion{O}{vi}).  Photospheric
\ion{Ne}{viii} lines are also identified in the hottest known He-rich
white dwarf (\kpd), some of which were also attributed to \ion{O}{viii}
previously.  This is a surprise because it must be concluded that \kpd\ is
much hotter (\Teff$\approx$200\,000~K) than hitherto assumed
(\Teff$\approx$120\,000~K). This is confirmed by a re-assessment of the
\ion{He}{ii} line spectrum. We speculate that the temperature is high
enough to explain the mysterious, hard X-ray emission (1~keV)
as being of photospheric origin.}


\maketitle
%

\begin{figure*}[tbp]
\begin{center}
\includegraphics[width=0.85\textwidth]{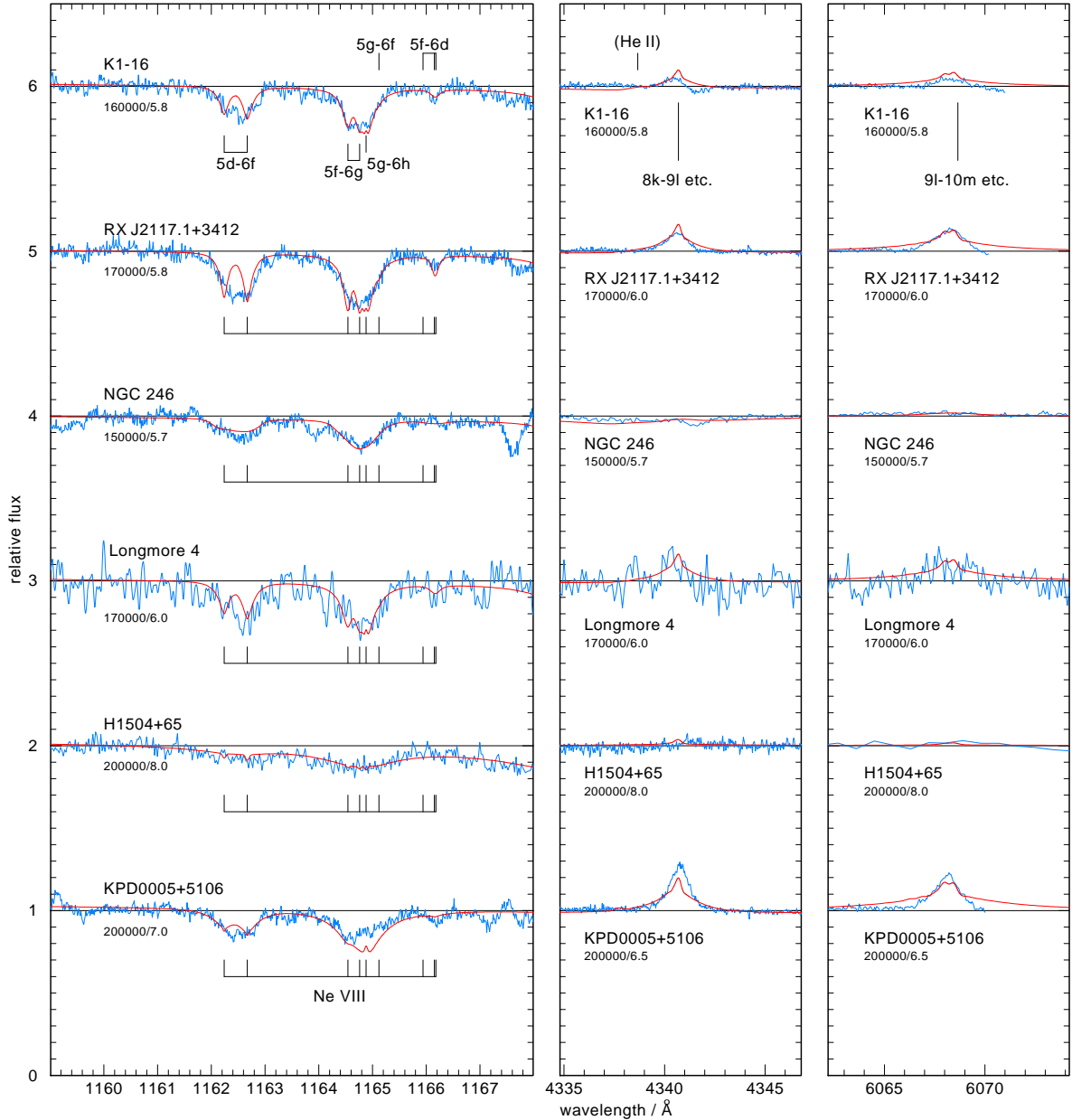}
  \caption[]{\emph{Left panel}: Identification of \ion{Ne}{viii} lines in the FUSE spectra
    of PG1159 stars and the DO white dwarf \kpd. Several lines of the
    $n=5 \rightarrow 6$ transition are detected, as labeled in detail at
    the uppermost spectrum. Overplotted are computed profiles with
    \Teff\ and $\log\,g$, as indicated. \emph{Middle and right
    panels}: Optical spectral regions where we identified the \ion{Ne}{viii}
    $n=8 \rightarrow 9$ and $n=9 \rightarrow 10$ transitions. All observed
    spectra are shifted such that the photospheric lines appear at their
    rest wavelengths.}
  \label{fig_all}
\end{center}
\end{figure*}

\begin{figure*}[tbp]
\begin{center}
\includegraphics[width=0.85\textwidth]{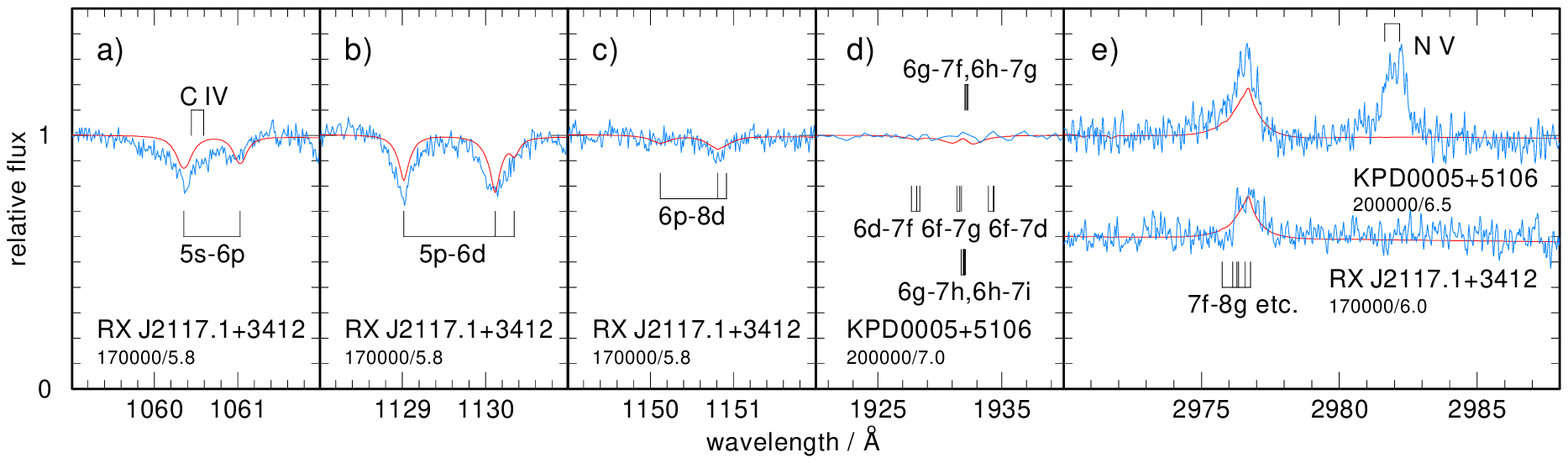}
  \caption[]{Details of other spectral regions of the PG1159-type
    central star \rxj\ and the DO white dwarf \kpd\ displaying  \ion{Ne}{viii} lines. \emph{Panels a)--c)} show
    $n=5 \rightarrow 6$ and $n=6 \rightarrow 8$ transitions (the blending
    \ion{C}{iv} line in panel \emph{a)} is not included in the model). \emph{Panel d)}
    shows a barely detectable $n=6 \rightarrow 7$ emission feature in the HST/FOS spectrum of \kpd, which
    was previously assigned to \ion{O}{viii}. \emph{Panel e)} displays HST/GHRS
    spectra with a \ion{Ne}{viii}  $n=7 \rightarrow 8$ emission line that
    was also thought to stem from \ion{O}{viii}. The adjacent emission
    feature in \kpd\ is from nitrogen; it is not included in the model.
}
  \label{fig_rxj2117_details}
\end{center}
\end{figure*}

\section{Introduction}
\label{intro}

Observations of extremely hot post-AGB stars (\Teff$\gappr 100\,000$~K)
with the \emph{Far Ultraviolet Spectroscopic Explorer} (FUSE) have led
to the discovery of many unidentified spectral lines. Their
identification is difficult, mostly because of the lack of accurate
atomic data for highly ionized elements. On the other hand, their
identification is rewarding because they are often the only features by
which particular species are accessible for abundance
determinations. Such species can be used to test stellar evolutionary
models. For example, several lines of fluorine (\ion{F}{v-vi}), neon
(\ion{Ne}{vii}), silicon (\ion{Si}{v}) and argon (\ion{Ar}{vii}) were
identified for the first time (Jahn \etal 2007; Werner \etal 2004a, 2005,
2007). In particular, the \ion{Ne}{vii} $\lambda~973$~\AA\ line occasionally
displays a prominent P~Cygni profile, which is an important
tool for stellar wind analyses (Herald \etal 2005).
In this paper, we report the identification of a number of
\ion{Ne}{viii} lines in the FUSE spectra of the hottest known
hydrogen-deficient (pre-) white dwarfs. We show that the mere occurrence
of such highly ionized neon lines puts a strict lower limit on the
effective temperature ($\approx 150\,000$~K).

\ion{Ne}{viii} is a one-valence electron, lithium-like ion with a
relatively simple  energy-term structure, however, a detailed
compilation of accurate level energies and line wavelengths, which is a
prerequisite of any quantitative work, became available only recently
(Kramida \& Buchet-Poulizac 2006). Our model atmosphere and spectrum
synthesis calculations predict the presence of \ion{Ne}{viii} lines in
other wavelength regions, too. As a surprising result, we find that all
previous identifications of ultra-high ionization (i.e., non-thermally
excited) oxygen lines (\ion{O}{viii}) in UV and optical spectra of
PG1159 stars, and the hottest known DO white dwarf, are wrong. We find
that, instead, these lines are due to photospheric \ion{Ne}{viii}. The
earlier identification of these lines as \ion{O}{viii} was motivated by
the supposed occurrence of the same lines in early-type Wolf-Rayet
central stars (i.e., spectral type [WCE]) and it was argued that they
are formed in shocked wind regions. The results presented here provide
evidence that also in [WCE] stars these, as well as other lines that were
assigned to ultrahigh-ionized C and O, do probably stem from thermally
excited neon.

We present our observations and line identifications in
Sect.\,\ref{observations} and describe our modeling in
Sect.\,\ref{modeling}. The results from line profile fits to individual
objects are presented in Sect.\,\ref{results}. Implications for [WCE]
stars are discussed in Sect.\,\ref{sect_wc} and we conclude with
Sect.\,\ref{conclusions}.

\begin{table}
\begin{center}
\caption{Effective temperature and surface gravity of the program stars as taken from the
literature. As described in the text, the discovery of \ion{Ne}{viii} lines
suggests that \Teff\ for \kpd\ and \lovier\ must be significantly higher. \label{tab_stars}}
\begin{tabular}{lcccc}
      \hline
      \hline
      \noalign{\smallskip}
Object  & Spectral &\Teff & \logg & Reference\\
        & Type     & [K]  & (cgs) &    \\
      \noalign{\smallskip}
      \hline
      \noalign{\smallskip}
\hh      & PG1159  & 200\,000 & 8.0 & A\\
\rxj     & PG1159  & 170\,000 & 6.0 & B\\
\ngc     & PG1159  & 150\,000 & 5.7 & C\\
\keins   & PG1159  & 140\,000 & 6.4 & D\\  
NGC\,2371& [WCE]   & 135\,000 & 5.5 & E\\
\lovier  & PG1159  & 120\,000 & 5.5 & F\\
\kpd     & DO      & 120\,000 & 7.0 & G\\
 \noalign{\smallskip}
      \hline
     \end{tabular}
\end{center}
References: 
A: Werner \etal 2004b,
B: Werner \etal 1996b,
C: Rauch \& Werner 1997,
D: Werner \etal 1992,
E: Herald \& Bianchi 2004,
F: Rauch \& Werner 1997,
G: Werner \etal 1994
\end{table}

\section{Observations and \ion{Ne}{viii} line identifications}
\label{observations}

FUSE observations and data reduction for most of our program stars
were described in our previous work (Werner \etal 2004a, 2004b, 2005,
2007). Table~\ref{tab_stars} lists our program stars with references to
results from previous analyses. The FUSE instrument consists of four
coaligned telescopes, each with a prime-focus spectrograph. Descriptions
of the FUSE instrument, and channel alignment and wavelength calibration
issues, are given by Moos \etal (2000) and Sahnow \etal (2000). The FUSE
spectra cover the wavelength range from the Lyman edge up to 1187~\AA\
with a spectral resolution of 0.05~\AA. For spectra that are too faint
to permit co-alignment of individual exposures, the spectral resolution
may be degraded to 0.1~\AA.  The S/N ratio in the LiF2a spectra in the
vicinity of the \ion{Ne}{viii} lines being studied here ranges from a
low of 17:1 per 0.05~\AA\ resolution element for \lovier\ to 70:1 per
resolution element for \rxj; it was greater than 45:1 for the other
objects.  The S/N ratio in the LiF1b spectra at these wavelengths was
typically $\sim$30\% lower.  UV spectra of \kpd\ and \rxj,
taken with the FOS and GHRS spectrographs aboard the \emph{Hubble Space
Telescope}, were retrieved from the MAST archive.  High-resolution
optical spectra of \keins, \rxj, \kpd, and \hh\ were obtained at the
\emph{Keck} observatory and the HIRES spectrograph. For details on data
reduction see Zuckerman \& Reid (1998). Spectra of \ngc\ and \lovier\
were obtained with the ESO \emph{Very Large Telescope} and the UVES
spectrograph, in the framework of the SPY project (Napiwotzki \etal 2003).

The starting point of our analysis was the identification of two strong
and broad absorption features of \ion{Ne}{viii} at
$\lambda=1162-1165$~\AA\ in the FUSE spectra of five PG1159 stars and
the DO white dwarf \kpd\ (left panel of Fig.\,\ref{fig_all}). They are
blends of several  $n=5\rightarrow 6$ lines with high angular quantum
number $l$. \ion{Ne}{viii} lines are not detected in the FUSE spectra of
any other PG1159 star, indicating that the minimum \Teff\ for exhibiting
these lines is around 150\,000~K (this limit will be assessed more closely
below). In addition to these lines, several weaker \ion{Ne}{viii} lines
are detectable in the FUSE spectra of some objects, particularly in
\rxj\ (panels \emph{a)--c)} in Fig.\,\ref{fig_rxj2117_details}). These
are low-$l$ $n=5\rightarrow 6$ lines and an isolated
$n=6\rightarrow 8$ line. The strongest, high-$l$,  $n=6\rightarrow 7$
transitions are located at $\lambda~1932$~\AA. We found only one
archival spectrum covering this wavelength position, namely an HST/FOS
observation of \kpd. An emission feature is barely detectable at this
location (panel \emph{d)} of Fig.\,\ref{fig_rxj2117_details}), 
formerly attributed to \ion{O}{viii} $n=6\rightarrow 7$ (Sion \&
Downes 1992). We note that the line positions for \ion{O}{viii} and
\ion{Ne}{viii} between levels with these high principal quantum numbers
become indistinguishable. Both ions have the same core charge but their
valence electrons are single $1s$ and $2s$ electrons,
respectively. Based on our model computations described below, we
attribute the previously discovered emission features at
$\lambda\lambda~2977, 4340$, and 6068~\AA\ (Figs.\,\ref{fig_rxj2117_details}
and \ref{fig_all}) to the $n=7\rightarrow 8$, $n=8\rightarrow 9$, and
$n=9\rightarrow 10$ transitions of \ion{Ne}{viii}, respectively. They,
too, were previously thought to stem from \ion{O}{viii} (Werner \& Heber
1992; Werner \etal 1996). The $\lambda~6068$~\AA\ emission line was also
detected in the very hot PG1159 stars Longmore~3 and HE\,1429$-$1209
(Werner \etal 1994, 2004a).

Table~\ref{tab} summarizes all \ion{Ne}{viii} lines that were detected in
any of the examined objects. The NIST\footnote{http://physics.nist.gov/
PhysRefData/ASD/index.html} wavelengths coincide with the photospheric
rest wavelengths, with the exception of the $5p\rightarrow 6d$,
$5f\rightarrow 6d$, and $6p\rightarrow 8d$ transitions. Their observed
wavelengths are smaller by 0.16, 0.09, and 0.07~\AA, respectively. This
is corrected for in panels \emph{b)} and \emph{c)} of
Fig.\,\ref{fig_rxj2117_details} and in Table~\ref{tab}.

We note that we have not detected any  \ion{Ne}{viii} lines in the
hottest (\Teff$\approx$100\,000~K) hydrogen-rich central stars or DA
white dwarfs from which FUSE spectra are available. The reason is that
their temperature is not sufficiently high (all have
\Teff$<$150\,000~K). The same holds for all hot DO white dwarfs besides
\kpd. The possible presence of \ion{Ne}{viii} lines in the FUSE spectra
of [WCE] stars is discussed below (Sect.\,\ref{sect_wc}).

\begin{table}
\begin{center}
\caption{\ion{Ne}{viii} lines identified in our program stars. 
Wavelengths are computed from NIST level energies (except for the $5p\rightarrow
6d$, $5f\rightarrow 6d$, and $6p\rightarrow 8d$ transitions, see text)
and given in vacuum or air for $\lambda$ smaller or larger than
3000~\AA, respectively. The presence of the feature at
$\lambda~1932$~\AA\ is uncertain.}
\label{tab} 
\begin{tabular}{ll}  
\hline 
\hline 
\noalign{\smallskip}
Wavelength / \AA &  Transition \\ \hline 
\noalign{\smallskip}
1060.36, 1061.03          &$ 5s\rightarrow 6p$\\
1129.02, 1130.12, 1130.35 &$ 5p\rightarrow 6d$ \\
1150.12, 1150.81, 1150.92 &$ 6p\rightarrow 8d$ \\
1162.24, 1162.67          &$ 5d\rightarrow 6f$ \\
1164.54, 1164.75          &$ 5f\rightarrow 6g$ \\
1164.88                   &$ 5g\rightarrow 6h$ \\
1165.94, 1166.15, 1166.18 &$ 5f\rightarrow 6d$ \\
1170.04, 1170.29          &$ 6d\rightarrow 8f$ \\
(1932.04)                 &$ 6h\rightarrow 7i$ etc.\\
2976.75                   &$ 7i\rightarrow 8k$ etc.\\
4340.77                   &$ 8k\rightarrow 9l$ etc.\\
6068.63                   &$ 9l\rightarrow 10m$ etc.\\
\noalign{\smallskip} \hline
\end{tabular} 
\end{center}
\end{table}

\section{Model atmospheres and neon line formation calculations}
\label{modeling}

We use a grid of line-blanketed non-LTE model atmospheres, which is
described in detail in Werner \etal (2004a). In essence, the models
include the main photospheric constituents, namely, He, C, O, and
Ne. The neon model atom consists of the ionization stages
\ion{Ne}{iv-ix}. Of particular importance for the investigations
presented in this paper is the \ion{Ne}{viii} ion. It consists of 5 NLTE
levels and 6 lines, which is sufficient for the atmospheric model
structure computation, but considerable extensions were required in order
to be able to compute profiles of the identified highly excited
lines. Since we aimed at the computation of lines involving the $n=10$
levels (the $n=9\rightarrow 10$ emission at 6068~\AA), we included all
77 levels up to $n=12$ with all 511 line transitions between them. In
addition, levels up to $n=14$ are included as LTE levels. With this
extended \ion{Ne}{viii} model ion, we performed line-formation iterations
for neon, i.e., keeping the atmospheric structure fixed. For the final
spectrum synthesis, fine-structure splitting of levels and lines was
considered, if possible. Energies for fine-structure splitting are not
available for all levels.  In particular, they are only partly available
for levels involved in the $n=5\rightarrow 6$ absorption lines at
$\lambda\ 1162-1165$~\AA. As a consequence, the detailed structure of
the computed line profile cores within these absorption troughs cannot
exactly match the observations and, indeed, a close inspection of
Fig.\,\ref{fig_all} reveals this problem. Level energies were taken from
NIST (Kramida \& Buchet-Poulizac 2006). Oscillator strengths were taken
from the Opacity (Seaton \etal 1994) and IRON (Hummer \etal 1993)
Projects databases (TIPTOPbase\footnote{
http://vizier.u-strasbg.fr/topbase/}), which are, however, not complete
for all transitions of our model atom. For lines involving levels with
$n=11$ and 12, we extrapolated oscillator strengths from transitions to
lower quantum numbers. For a lithium-like ion this appears reasonably
accurate; the estimates are probably within a 10\%
error. Photoionization cross-sections are taken from the Opacity Project
database when available or, otherwise, computed in hydrogen-like
approximation. Electron collisional rates were calculated with usual
approximation formulae. The Ne model atoms that were used for this
analysis have been developed in the framework of the German
Astrophysical Virtual Observatory
(\emph{GAVO}\footnote{http://www.g-vo.org}) project and are provided
within the T\"ubingen Model-Atom Database
\emph{TMAD}\footnote{http://astro.uni-tuebingen.de/\raisebox{.2em}{\tiny
$\sim$}rauch/TMAD/TMAD.html}.

The treatment of Stark broadening of \ion{Ne}{viii} lines poses a severe
problem. We employ an approximate formula to compute Stark widths, which
we use routinely to compute profiles of lines between highly excited
levels of lithium-like ions (\ion{C}{iv}, \ion{O}{vi}; Werner \etal
1991), and which accounts for linear Stark broadening. While this
formula gave reasonable results for the other Li-isoelectronic ions in
our past work, we found that the $n=5\rightarrow 6$ absorption lines in
the FUSE spectral  range came out much too narrow (by about a factor of
three) compared to all of  the observations. Interestingly, a similar
notorious problem is encountered in the interpretation of \ion{Ne}{viii}
line widths of laboratory spectra. Measurements of the $3s\rightarrow
3p$ transition at 2821~\AA\ (Glenzer \etal 1992) display line profiles
that are roughly a factor of two wider than predicted by theory, i.e.,
the $Z^{-2}$ scaling of the line width, as predicted by the impact theory,
significantly deviates from observations. Even today, this discrepancy
is not resolved, despite of intense efforts with both improved
experiments and quantum-mechanical calculations (Hegazy \etal 2003;
Griem \& Ralchenko 2006). Our finding from stellar spectra indirectly
suggests that the line widths measured in laboratory are in fact correct,
and that instead, the line-broadening theory is still inaccurate. It is
necessary to note, however, that we observe different lines of \ion{Ne}{viii}.

To cope with this problem we formally reduced the ionic core charge of
\ion{Ne}{viii} in our line broadening formula from $Z=8$ to $Z=2$. This
yields widths of the $n=5\rightarrow 6$ lines that coincide with the
observations. Of course, this procedure is rather unsatisfactory and
would prohibit an accurate abundance determination from these lines. In addition,
a close inspection of the computed profiles for the optical lines (i.e.,
the higher-$n$ transitions, Fig.\,\ref{fig_all}) reveals that they
become too broad with our crude $Z$-reduction. Fortunately, the neon
abundance of all PG1159 stars discussed here is known to be of the order
2\% (Werner \etal 2004a). The focus of our paper is on the extreme
temperature sensitivity of the \ion{Ne}{viii} lines and line broadening
is a subordinate effect only.

\begin{figure}[tbp]
\begin{center}
\includegraphics[width=0.95\columnwidth]{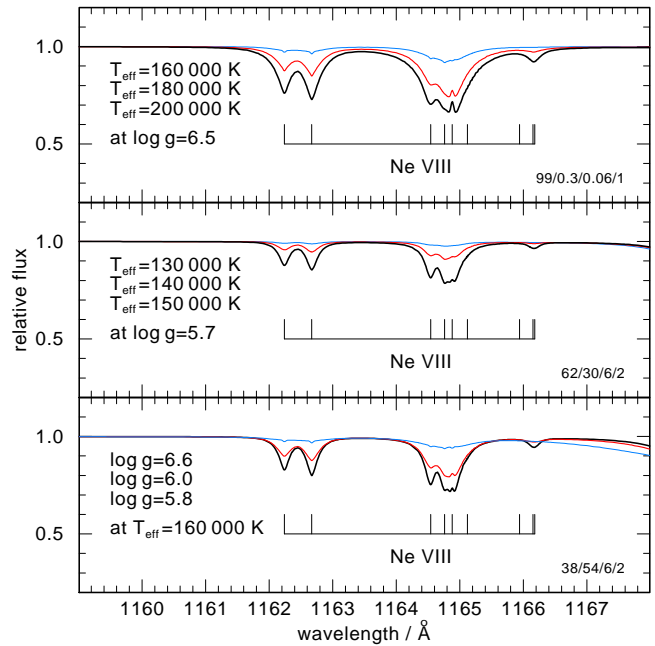}
  \caption[]{The \ion{Ne}{viii} lines are strongly sensitive to
    \Teff\ and $\log\,g$. Generally, they become stronger with increasing \Teff\ and
    decreasing $\log\,g$.
\emph{Top panel}: \Teff-sensitivity in DO models with \logg=6.5. The
lines disappear if \Teff\ is below 160\,000~K.
\emph{Middle panel}: \Teff-sensitivity in PG1159 models with lower
gravity (\logg=5.7). In this case, the lines are detectable down to \Teff=140\,000~K.
\emph{Bottom panel}: \logg\ sensitivity in PG1159 models at
\Teff=160\,000~K. The gravity must be sufficiently low, otherwise the
\ion{Ne}{viii} lines are not detectable. The numbers in the lower right
corner of the panels give the model abundances of He/C/O/Ne in \% mass fraction.}
  \label{fig_teff_variation}
\end{center}
\end{figure}

\begin{figure*}[tbp]
\begin{center}
\includegraphics[width=0.85\textwidth]{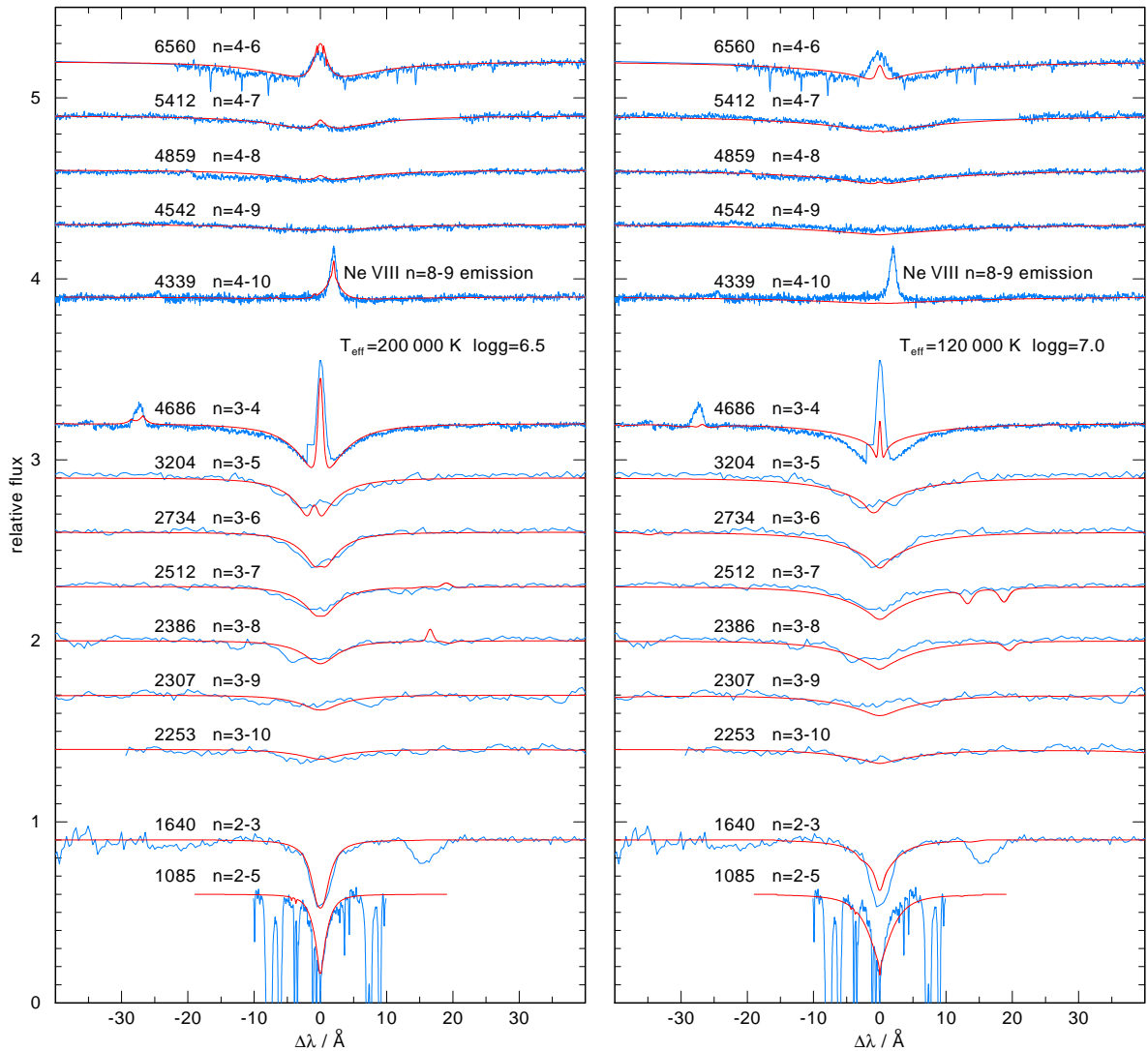}
  \caption[]{The \ion{He}{ii} lines in \kpd\ are compared with two different
    models. \emph{Left panel}: \Teff=200\,000~K, \logg=6.5. \emph{Right panel}:
    \Teff=120\,000~K, \logg=7.0. The hotter model yields a much better
    fit. In particular, only this model matches the emission cores in the $\lambda\lambda~6560$ and $4686$~\AA\
    lines. The emission line at $\lambda~4340$~\AA\ is due to \ion{Ne}{viii}
    and is only exhibited in the hotter model.}
  \label{fig_kpd0005_fit}
\end{center}
\end{figure*}

\section{Results}
\label{results}

We now compare our computed \ion{Ne}{viii} line profiles to the
observations (Fig.\,\ref{fig_all}). It is instructive to look at
Fig.\,\ref{fig_teff_variation}, in which we demonstrate that the
strengths of the $\lambda\lambda~1162-1165$~\AA\ absorption lines
strongly depend on \Teff\ and $\log\,g$.  The minimum \Teff\ for a line
detection is a strong function of gravity, ranging from about 160\,000~K
at \logg=6.5 to 140\,000~K at \logg=5.7.  The mere detection of these
\ion{Ne}{viii} lines thus provides a strict lower limit to $T\mathrm{\hspace*{-0.4ex}_{eff}}$.
Generally, the overall characteristics of the \ion{Ne}{viii} lines are
reproduced by the models. We find that the UV lines are in absorption,
while the optical lines are in emission.

Let us discuss in some detail each of the six objects in which we
detected the UV \ion{Ne}{viii} lines, particularly those cases in which
contradictions to \Teff\ determinations from previous work appear. This
mainly affects the DO white dwarf \kpd, whereas for the PG1159 stars the
\ion{Ne}{viii} line features can be matched with model parameters that
are in good or acceptable agreement with previous results.

\subsection{PG1159 stars}

Models for individual objects were computed with element abundances as
given in Werner \etal (2005). As already mentioned, the neon abundance
is kept fixed at 2\%. We started with values for \Teff\ and \logg\ also
taken from this reference; typical uncertainties are 10\% and 0.5~dex,
respectively.

\paragraph{\keins} The previously determined parameters are
\Teff=140\,000~K and \logg=6.4. Clearly, a model with these values shows
no \ion{Ne}{viii} lines at all. We find an acceptable fit  only at
\Teff\ and \logg\ values that are slightly beyond the error bars, namely
\Teff=160\,000~K and \logg=5.8.

\paragraph{\rxj} We find a good fit to the \ion{Ne}{viii} lines with
models close to the literature values (\Teff=170\,000~K, \logg=6.0). At
these parameter values, the \ion{Ne}{viii} absorption lines are
strongest. This explains why we can identify a number of additional weak
\ion{Ne}{viii} lines in the FUSE spectrum
(Fig.\,\ref{fig_rxj2117_details}).

\paragraph{\ngc} A good fit to the \ion{Ne}{viii} lines is
obtained at the literature values (\Teff=150\,000~K, \logg=5.7). Note
that \ngc\ is a fast rotator and the computed profiles were broadened
with 70~km/s (Rauch \& Werner 1997).

\paragraph{\lovier} The strong \ion{Ne}{viii} lines in the FUSE spectrum
suggest that this star is as hot as \rxj. We find a good fit with a
\Teff=170\,000~K, \logg=6.0 model. Therefore, \lovier\ is significantly
hotter than previously thought (\Teff=120\,000~K, \logg=5.5). This
result is, however, not surprising because we have already found
independent hints that the temperature is at least about 150\,000~K
(Werner \etal 2004a). Note that \lovier\ exhibited much more prominent
optical \ion{Ne}{viii} lines during its observed ``outburst'' and change
of spectral type from PG1159 to [WCE] (Werner \etal 1992).

\paragraph{\hh} The \ion{Ne}{viii} lines in the FUSE spectrum are
shallow and broad. This is a consequence of the high gravity (8.0). The
profiles confirm the extremely high \Teff\ (200\,000~K).

\subsection{The DO white dwarf \kpd}

An analysis of optical and HST UV spectra of \kpd\ gave the result
\Teff=120\,000~K and \logg=7 (Werner \etal 1994). Although being the
hottest known DO white dwarf, the identification of \ion{Ne}{viii} lines
is a big surprise. Figure~\ref{fig_teff_variation} shows that at this
relatively high gravity the temperature must be around 180\,000~K. We
find that models with \Teff=180\,000--200\,000~K and \logg=6.5--7.0 give
the best fit to the \ion{Ne}{viii} lines. Interestingly, this is not in
contradiction with the \ion{He}{ii} lines. We even find that a model
with \Teff=200\,000~K and \logg=6.5 gives better \ion{He}{ii} line fits
than the cooler 120\,000~K model. This is shown in
Fig.\,\ref{fig_kpd0005_fit}. To be more specific, the hot model is able
to match the height of the central emissions reversals in the
$\lambda\lambda~4686$ and $6560~$\AA\ line cores. At the same time, the
$\lambda~1640$~\AA\ line is matched very well.  We conclude that \kpd\
is significantly hotter than previously thought.

Unlike for the PG1159 stars, the neon abundance in \kpd\ is not known
from previous analyses. The usual diagnostic lines are \ion{Ne}{vii}
$\lambda\lambda~973$ and $3644~$\AA. The FUSE spectrum of \kpd\ is
contaminated by interstellar H$_2$ absorption, making the detection of \ion{Ne}{vii}
$\lambda~973~$\AA\ impossible. In a medium-resolution optical spectrum
published by Werner \etal (1994), no line feature is detected at
$\lambda~3644~$\AA. We have verified that this is compatible with all
our models presented here in the \Teff=120\,000-200\,000~K range.

We assumed that the Ne abundance is determined
by radiative levitation. The hottest model presented by Chayer \etal
(1995) has \Teff=100\,000~K with \logg=7.5 and predicts $\log({\rm
Ne/He})=-4$ by number. It is difficult to extrapolate this to a significantly
higher \Teff\ but the tendency is that the Ne abundance increases with
$T\mathrm{\hspace*{-0.4ex}_{eff}}$. A further increase can be expected because of the smaller
gravity. Although uncertain, it is not unreasonable that we set Ne=1\%
(by mass), that is, $\log({\rm Ne/He})=-2.7$ by number. We stress that
the high \Teff\ derived from the \ion{Ne}{viii} lines is hardly affected
by this assumption. In addition, the \ion{He}{ii} line profiles are
essentially independent of the neon abundance.

\section{Implications for ultrahigh-ionisation emission line identifications in [WCE] stars}\label{sect_wc}

Since the supposed identification of several optical emission lines in
 the hot [WC] star \sand\ as being due to ultrahigh ionised
 (i.e. non-thermally excited) C and O (\ion{C}{v} and \ion{O}{vii-viii})
 (Barlow \etal 1980), a number of authors claimed the identification of
 these features in other early-type [WC] stars. In the light of
 our results, we propose that most, and probably all, of these features
 stem from thermally excited ionisation stages of neon
 (\ion{Ne}{vii-viii}), carbon and oxygen. Table~\ref{tab2} summarizes
 our proposed identifications.  We give qualitative arguments for this,
 but detailed NLTE modeling with expanding model atmospheres will be
 necessary for a quantitative confirmation.

\begin{figure}[tbp]
\begin{center}
\includegraphics[width=0.95\columnwidth]{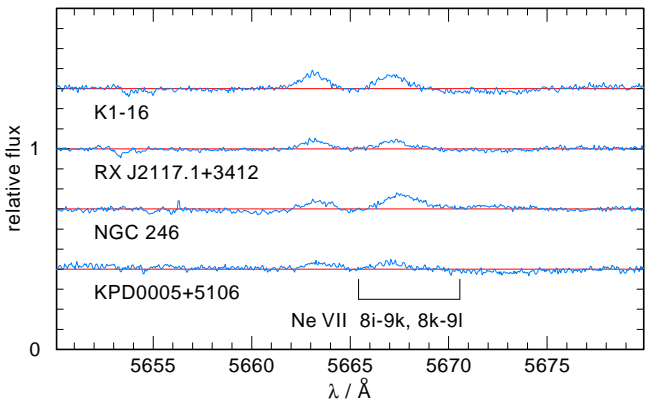}
  \caption[]{First identification of two high-$l$ $n=8\rightarrow 9$ emission
    lines of \ion{Ne}{vii} (!). A strong and broad emission feature is seen
    in [WCE] stars at this location. We propose that it is due to these
    \ion{Ne}{vii} transitions and not due to ultrahigh ionized
    \ion{O}{vii}. The labels denote the theoretical line positions using
    energy levels that are not precisely known.}
  \label{ne7_5660}
\end{center}
\end{figure}

A broad emission feature in \sand\ and other [WCE]s at
$\lambda~3893$~\AA\ was attributed to the $n=7 \rightarrow 8$ transition
of \ion{O}{vii}. We propose that it originates from a prominent
\ion{Ne}{vii} multiplet recently identified in the hottest PG1159
central stars (Werner \etal 2004a), which has its strongest component
located at $\lambda~3892$~\AA. In addition, the \ion{Ne}{vii} $n=7
\rightarrow 8$ transition contributes. The emission at
$\lambda~4340$~\AA\ is probably from \ion{Ne}{viii}, not
\ion{O}{viii}. Emission lines at $\lambda\lambda~4555$~\AA\ and
5665~\AA\ were assigned to the \ion{O}{vii} transitions $n=9 \rightarrow
11$ and $n=8 \rightarrow 9$, respectively. We suggest that they are
lines between high Rydberg states of \ion{Ne}{vii}, with the same
principal quantum numbers (level energies are given in NIST and Lapierre
\& Knystautas 1999). This is supported by our discovery of two high-$l$
$n=8\rightarrow 9$ emission lines of \ion{Ne}{vii} at
$\lambda~5665$~\AA\ in some of our program stars (Fig.\,\ref{ne7_5660}).
Finally, the strong emission feature at $\lambda~4945$~\AA\ was assigned
to \ion{C}{v} $n=7 \rightarrow 6$. We think that it is the respective
\ion{N}{v} transition, as in the case of \kpd. Nitrogen is definitely
present in \sand, because the $3s\rightarrow 3p$ emission doublet at
$\lambda\lambda~4602, 4620$~\AA\ is present, as well as a prominent
P~Cygni profile of the \ion{N}{V} resonance line. Thus, the entire
optical emission line spectrum of \sand\ and other [WCE]s can be
explained without invoking ultrahigh-ionisation features.

\begin{table}
\begin{center}
\caption{List of lines for which we propose photospheric identifications
    as opposed 
  to previously thought identifications as non-photospheric
  ultrahigh-ionisation features in PG1159 stars, [WCE] stars, and in the
  DO \kpd. The \ion{O}{vi} and  \ion{N}{v} emission lines were
  identified before but thought to be blended by the ultrahigh ionisation
  features.} 
\label{tab2} 
\begin{tabular}{lll}  
\hline 
\hline 
\noalign{\smallskip}
Wavelength / \AA & Old ultrahigh-    & New photospheric \\
                 & ionisation        & identification   \\
                 & identification    &                  \\ \hline 
\noalign{\smallskip}
1932             & \ion{O}{viii} \quad $n=7\rightarrow  8$ & \ion{Ne}{viii} \quad $n=7\rightarrow 8$ \\
2977             & \ion{O}{viii} \quad $n=6\rightarrow  7$ & \ion{Ne}{viii} \quad $n=6\rightarrow 7$ \\
3893             & \ion{O}{vii}  \quad $n=7\rightarrow  8$ & \ion{Ne}{vii} 3p\,$^3$P$^{\rm o}\rightarrow$ 3d\,$^3$D\\
                 &                                   & plus \ion{Ne}{vii} \quad $n=7\rightarrow  8$ \\
4340             & \ion{O}{viii} \quad $n=8\rightarrow  9$ & \ion{Ne}{viii} \quad $n=8\rightarrow 9$ \\
4500             & \ion{C}{vi}   \quad $n=8\rightarrow 10$ & \ion{O}{vi}    \quad $n=8\rightarrow 10$ \\
4555             & \ion{O}{vii}  \quad $n=9\rightarrow 11$ & \ion{Ne}{vii}  \quad $n=9\rightarrow 11$ \\
4945             & \ion{C}{v}    \quad $n=6\rightarrow  7$ & \ion{N}{v}     \quad $n=6\rightarrow  7$ \\  
5290             & \ion{C}{vi}   \quad $n=7\rightarrow  8$ & \ion{O}{vi}    \quad $n=7\rightarrow 8$ \\
5665             & \ion{O}{vii}  \quad $n=8\rightarrow  9$ & \ion{Ne}{vii}  \quad $n=8\rightarrow 9$  \\
6068             & \ion{O}{viii} \quad $n=9\rightarrow 10$ & \ion{Ne}{viii} \quad $n=9\rightarrow 10$ \\
\noalign{\smallskip} \hline
\end{tabular} 
\end{center}
\end{table}

Based on observations  with the \emph{International Ultraviolet
Explorer} (IUE), several papers appeared in the literature claiming the
presence of ultrahigh ionisation lines in the UV spectra of [WCE]
stars. In \sand, as well as in NGC~5315 and NGC~6905, Feibelman (1996a,
b) discovered the $\lambda\lambda~1932$~\AA\ and 2977~\AA\ emission
features that we have discussed above, and he also assigned them to
\ion{O}{viii}. As in \kpd\ and \rxj\
(Fig.\,\ref{fig_rxj2117_details}), they probably are from \ion{Ne}{viii}.

\begin{figure}[tbp]
\begin{center}
\includegraphics[width=0.95\columnwidth]{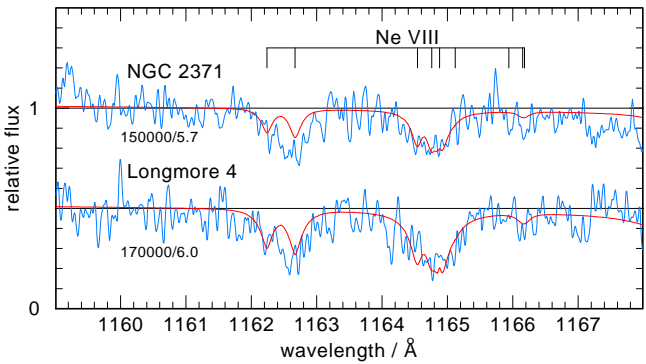}
  \caption[]{Identification of \ion{Ne}{viii} lines in the [WCE] central
    star NGC~2371 (\emph{top}) and comparison to the PG1159-type central star
    \lovier\ (\emph{bottom}).}
  \label{wcstars}
\end{center}
\end{figure}

Feibelman (1996b) even claimed the existence of an ``\ion{O}{viii}
sequence'' of planetary-nebulae nuclei, mainly based on the supposed
\ion{O}{vii-viii} identifications in PG1159 and [WCE] stars. Obviously,
that idea must now be discarded. This is further corroborated by the
fact that we detected the \ion{Ne}{viii} lines at
$\lambda~1162-1165$~\AA\ in a [WCE] star too, namely in NGC~2371
(Fig.\,\ref{wcstars}). The line strengths suggest \Teff$\approx
150\,000$~K, which is, considering error limits, in good agreement with
the result of a detailed analysis of FUSE and IUE spectra (135\,000~K;
Herald \& Bianchi 2004). NGC~2371 was classified as [WC4] by Acker \&
Neiner (2003). These authors have based a new spectral classification
system for the hottest [WC]s on the occurrence of \ion{O}{vii-viii}
lines in the optical spectrum. Our identification as \ion{Ne}{vii-viii}
lines means that the empirical classification criteria remain
essentially correct because only the hottest [WC]s are able to show
these lines.

\section{Summary and conclusions}
\label{conclusions} 

We have identified \ion{Ne}{viii} absorption lines in FUSE spectra of
PG1159 stars and a DO white dwarf. Line profile fits  confirm that the
PG1159 stars are the very hottest members of their spectral class
(\Teff$\geq$150\,000\,K). We have shown that two well-known emission
lines in the optical spectra of these stars are also from \ion{Ne}{viii}
and not, as previously thought, from \ion{O}{viii}. While \ion{Ne}{viii}
is thermally excited in the hot photospheres, the existence of
\ion{O}{viii} would require temperatures of the order of one million~K,
thus, an unknown process (e.g., shock zones in the wind) was invoked to
explain these emission lines. This is no longer necessary.

We argue that these, and probably all, emission lines in early-type [WC]
central stars that are usually assigned to ultrahigh-ionisation stages
(\ion{O}{vii-viii}, \ion{C}{v}) originate from \ion{Ne}{vii-viii} and
other thermally excited species.

The discovery of \ion{Ne}{viii} lines in the central star \lovier\
suggests that its \Teff\ is distinctively higher than previously thought
(170\,000~K instead of 120\,000~K). To a smaller extent, the same
tendency is seen for the PG1159 star \keins\ and the [WCE] star
NGC\,2371. This calls for a re-analysis of the complete spectra of these
stars to see how the fit to other line features can be reconciled. The
computation of model grids with varying parameters $T\mathrm{\hspace*{-0.4ex}_{eff}}$, $\log\,g$, and
abundances of the most important elements (He, C, O, Ne) will be
necessary. Since the stars display prominent P~Cygni line profiles,
it could be advantageous to use expanding atmosphere models with the
available spectral data. 
As already mentioned, the neon abundance cannot
be determined from the newly discovered \ion{Ne}{viii} lines because of
the uncertainties with the line broadening theory. Instead, the strong
P~Cygni profile of the \ion{Ne}{vii} $\lambda~973~$\AA\ line exhibited
by these stars can be used as a sensitive tool (Bianchi \& Herald
2007). Again, detailed parameter studies must be performed to find
which neon abundance and which mass-loss rate yield good fits to this
\ion{Ne}{vii} line profile, while at the same time determining the \Teff\ of the model that
is high enough to produce detectable \ion{Ne}{viii} lines.

A surprising result of our investigation is the identification of
\ion{Ne}{viii} lines in the hottest known DO white dwarf, \kpd. We
conclude that its temperature is close to \Teff=200\,000~K and, hence,
that it is significantly hotter than hitherto thought (120\,000~K). As
in the case of the PG1159 stars, the optical \ion{Ne}{viii} emission
lines were previously assigned to ultrahigh-ionized \ion{O}{viii}. The announcement of the discovery of a relatively \emph{soft}
X-ray emitting corona about \kpd\ seemed to support this assignment
(Fleming \etal 1993). However, the analysis of a \emph{Chandra} spectrum
exhibiting flux in the range 20--80~\AA\ proved that the soft X-rays
stem from the photosphere of \kpd\ (Drake \& Werner 2005). The
deposition of the corona is now in accordance with the deposition of the
ultrahigh-ionisation lines. A comprehensive
re-analysis of all available data (\emph{Chandra}, FUSE, HST,
\emph{Keck}) is required to tightly constrain the atmospheric parameters and metal
abundances.

It remains to be seen how the extremely high effective temperature
derived in our work relates to the observed \emph{hard} X-ray emission
at 1~keV (12~\AA) from \kpd\ (O'Dwyer \etal 2003). It might be that this
emission is also of photospheric origin. After all, the entire spectral
properties of \kpd\ could be deciphered as thermal photospheric
radiation. In any case, the star has turned out to be \emph{by far} the
hottest known DO white dwarf.

\begin{acknowledgements}
We are grateful to the referee, Luciana Bianchi, for a careful reading
of the manuscript and for her constructive comments.
We thank Uli Heber and Ralf Napiwotzki for putting their \emph{Keck} and
\emph{VLT} spectra at our disposal.  We thank A.E.~Kramida, H.R.~Griem,
\mbox{H.-J.} Kunze, M.S.~Dimitrijevi{\'c}, and {\'E}. Knystautas for
discussions on atomic data and sending us reprints. T.R. is supported
by the \emph{German Astrophysical Virtual Observatory} project of the
German Federal Ministry of Education and Research under grant
05\,AC6VTB. J.W.K. is supported by the FUSE project, funded by NASA
contract NAS5-32985.
\end{acknowledgements}

\end{document}